\documentstyle[sprocl,psfig]{article}

\def\be{\begin{equation}}
\def\ee{\end{equation}}
\def\bea{\begin{eqnarray}}
\def\eea{\end{eqnarray}}
\begin{document}

\title{VELOCITY PEAKS AND CAUSTIC RINGS} 

\author{P. SIKIVIE}

\address{Department of Physics, University of Florida,\\
Gainesville, FL 32611 USA\\E-mail: sikivie@phys.ufl.edu}

\maketitle\abstracts{The late infall of cold dark matter onto 
an isolated galaxy produces flows with definite local velocity 
vectors throughout the galactic halo.  It also produces caustic 
rings, which are places in the halo where the dark matter density 
is very large.  The self-similar model of halo formation predicts 
that the caustic ring radii $a_n$ follow the approximate law 
$a_n \simeq 1/n$.  I interpret bumps in the rotation curves of 
NGC 3198 and of our own galaxy as due to caustic rings of dark 
matter.  In this model of our halo the annual modulation effect 
in direct searches for WIMPs has the opposite sign from that
predicted by the isothermal sphere model.}

\vspace{0.5 cm}
 
\section{Late infall}

There are compelling reasons to believe that the dominant component of 
the dark matter of the Universe is non-baryonic collisionless particles,
such as axions, weakly interacting massive particles (WIMPs) and massive
neutrinos. The word ``collisionless'' signifies that the particles are 
so weakly interacting that they have moved purely under the influence of 
gravity since their decoupling at a very early time (of order $10^{-4}$ 
sec for axions, of order 1 sec for neutrinos and WIMPs).  In the limit 
where the primordial velocity dispersion of the particles is neglected, 
they all lie on the same 3-dimensional (D) `sheet' in 6-D phase-space.
Their phase-space evolution must obey Liouville's theorem.  This implies 
that the sheet cannot tear and hence that it satisfies certain topological 
constraints.

Because their phase-space sheet cannot tear, collisionless dark matter
(CDM) particles must be present everywhere in space, including specifically 
intergalactic space.  The space density may be reduced by stretching of the 
sheet, but it cannot vanish.  Moreover, the average space density is 
recovered as soon as the average is taken over distances larger than 
the distance CDM may have moved locally away from perfect Hubble flow.  
In a region that is sparsely populated with galaxies, this distance 
is much smaller than the distance between galaxies.  The implication is that 
isolated galaxies are surrounded by unseen CDM and hence, because of gravity, 
CDM keeps falling continuously onto such galaxies from all directions.  
If the galaxy joins other galaxies to form a cluster, infall onto the galaxy 
gets shut off because of lack of material, but infall onto the cluster 
continues assuming that the cluster is itself isolated.  In an open 
universe $(\Omega < 1)$, the infall process eventually turns off because 
the universe becomes very dilute.  However, even if our universe is open, 
we are far from having reached the turn-off time.  

\section{Velocity peaks}

Consider the infall of CDM onto an isolated galaxy.  Let us at first
neglect the velocity dispersion of the infalling particles.  In practice
it is sufficient that their velocity dispersion is much smaller than 
the rotation velocity of the galaxy.  Consider the time evolution of all 
CDM particles that are about to fall onto the galaxy for the first time 
in their history at time $t$.  For the sake of definiteness, we may consider 
all particles which have zero radial velocity $(\dot{r} =0)$ for the first 
time then.  Such particles are said to be at their `first turnaround';
they were receding from the galaxy as part of the general Hubble flow 
before $t$ and will be falling onto the galaxy just after $t$.  They form 
a closed surface, enclosing the galaxy, called the turnaround 'sphere' at 
time $t$.  The present turnaround sphere of the Milky Way galaxy has a 
radius of order 2 Mpc.  The turnaround sphere at time $t$ falls through 
the central parts of the galaxy at a time of order $2t$.  Particles falling 
through the galactic disk (assuming the galaxy is a spiral) get scattered 
through an angle $\Delta \theta \sim 10^{-3}$ by the gravitational fields
of various inhomogeneities such as molecular clouds, globular clusters, 
and stars \cite{is}. However, most particles carry too much angular momentum
to reach the luminous parts of the galaxy and are scattered much less.  
Because the scattering is small, the particles on the turnaround sphere 
at time $t$, after falling through the galaxy, form a new sphere which 
reaches its maximum radius $R'$ at some time $t'$.  The radius $R'$ at 
the second turnaround is smaller than the radius $R$ at the first
turnaround because the galaxy has grown by infall in the meantime.  The 
sphere continues oscillating in this way although it gets progressively 
fuzzier because of scattering off inhomogeneities in the galaxy.

\begin{figure}[top]
\psfig{figure=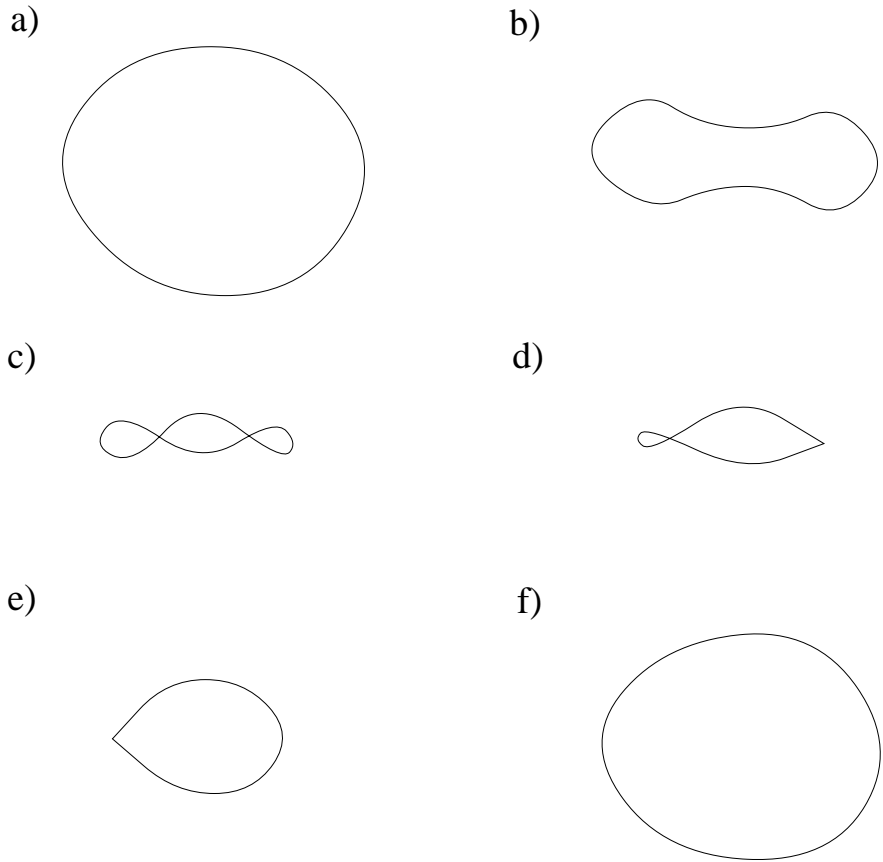,height=4.5in}
\vskip 2cm
\caption{\small{Infall of a turnaround sphere.  The sphere has net 
angular momentum about the vertical axis.  It crosses itself between 
frames b) and c).  After frame e) the sphere has completed the process
of turning itself inside out.  The cusps in frames d) and e) are at 
the intersection of a ring caustic with the plane of the figure.}}
\label{fig1}
\end{figure}

Fig.~1 describes the fall of one turnaround sphere through the galaxy
as a succession of time frames.  No particular symmetry is assumed.  It
is assumed, for the sake of definiteness, that the particles on the 
turnaround sphere carry net angular momentum about the vertical axis. 
The particles which are near the top in frame a) carry little angular 
momentum.  They fall through the center of the galaxy and end up near the 
bottom of the turnaround sphere in frame f).  Similarly the particles 
which are near the bottom in frame a) end up near the top in frame f).
The particles which are near the equator carry the most angular momentum.  
They form a ring whose size decreases to some minimum value and then 
increases again.  In the process of falling through the galaxy the 
turnaround sphere turns itself inside out.  Note also that the sphere 
crosses at least twice each point which is inside the sphere both at 
the initial time of frame a) and at the final time of frame f).  

Fig.~1 shows the motion of just one sphere in a continous flow of such 
spheres.  Moreover, there are many flows in and out of the galaxy going 
on at the same time.  To each flow in and out of the galaxy is associated 
a pair of peaks in the velocity distribution of CDM particles at every 
physical point in the galactic halo \cite{is}.  One peak is due to 
particles falling onto the galaxy for the first time, one peak is due to 
particles falling out of the galaxy for the first time, one peak is due 
to particles falling onto the galaxy for the second time, and so on.  In 
particular this is true on Earth.  Igor Tkachev, Yun Wang and I \cite{twi} 
obtained estimates of the velocity magnitudes and the local density 
fractions associated with these peaks using the self-similar infall model 
of galactic halo formation.  We generalized the existing version of the 
model \cite{ssi} to take account of the angular momentum of the dark matter 
particles.  We find that the first twenty peaks contribute each between 
one to four percent of the local density.  This implies a large 
non-thermal component to the galactic CDM distribution.

\section{Caustic rings}

There is a caustic ring associated with each flow in and out of the 
galaxy \cite{ring}.  A caustic is a place in physical space where the 
density is large because the phase-space sheet `folds back' there.  At 
the caustic, the space density diverges in the limit of zero velocity 
dispersion.  In Fig.~1, the caustic ring intersects the plane of the 
figure at the location of the cusps in frames d) and e).  One can prove 
mathematically \cite{tbp} that the density diverges at these points in
the limit of zero velocity dispersion of the infalling particles.  In 
reality the infalling particles have some velocity dispersion.  However 
only when this velocity dispersion is as large as 30 km/s do the caustic 
rings in our galaxy get washed out.  It is generally believed that the 
velocity dispersion of infalling CDM particles is at most 10 km/s.

For an arbitrary angular momentum distribution on the turnaround sphere, 
the ring is a closed loop of arbitrary shape.  However, if the angular 
momentum distribution is dominated by a smooth component that carries 
\emph{net} angular momentum, the ring resembles a circle.  If there is 
no angular momentum at all, the ring reduces to a point at the galactic 
center.  

The caustic ring is located in physical space near where the particles 
with the largest amount of angular momentum are at their distance 
of closest approach to the galactic center.  In the particular case where 
the turnaround sphere is initially rotating with a definite angular 
velocity as if it were a rigid body, the density distribution near the 
caustic is:
\begin{equation}
d(a;\rho,z) \simeq {dM \over d\Omega dt}~ {2 \over v}~
{1 \over \sqrt{(r^2-a^2)^2 + 4 a^2 z^2}}
\end{equation}
where $(\rho,z,\theta)$ are cylindrical coordinates, $a$ is the caustic
ring radius assumed to be much smaller than $R$, $r= \sqrt{\rho^2 + z^2},
~{dM \over d\Omega dt}$ is the rate at which mass falls in per unit time 
and unit solid angle, and $v$ is the velocity magnitude of the particles
at the caustic.  Near the caustic, the density diverges as the inverse
distance to the ring:
\begin{equation}
d(a;\rho,z) \simeq {dM\over d\Omega dt}~ {1\over v~ a~ \sigma}
\label{(3)}
\end{equation}
with $\sigma = \sqrt{(\rho -a)^2 + z^2}$.

The self-similar infall model mentioned earlier \cite{twi} predicts the
values of successive radii to be \cite{ring}:
\begin{equation}
\{a_n: n=1,2,3...\} \simeq (39,19.5,13,10,8 ...) {\rm kpc}
\left({j_{max} \over 0.25}\right) \left({0.7 \over h}\right)
\left({v_{rot} \over 220 {\rm km/s}}\right)
\end{equation}
for $\epsilon=0.3$.  $\epsilon$ is a parameter which is predicted by 
theories of large scale structure formation to be in the range 0.2 to
0.35, $j_{max}$ is the maximum value of the angular momentum of the
particles on the turnaround sphere in the dimensionless units 
defined in Ref.[2], $h$ is the Hubble rate in units of 
100 km/sec.Mpc and $v_{rot}$ is the rotation velocity of the galaxy.
For $\epsilon = 0.2,~ a_1 \simeq 36 {\rm kpc}~\left({j_{max}\over 0.25}
\right) \left({0.7\over h}\right) \left({v_{rot}\over 220
{\rm km/s}}\right)$, but the ratios $a_n/a_1$ are almost the same as 
in the $\epsilon = 0.3$ case.  Thus in the range of interest $0.2 \leq
\epsilon \leq 0.35$, the ratios of ring radii are nearly $\epsilon$ 
independent and follow the approximate law: $a_n \sim 1/n$.  The 
self-similar model also predicts for each ring the prefactor 
${dM \over d\Omega dt}~ {2 \over v}$ that appears in Eq.(1).  For 
example, for $\epsilon = 0.2$, 
$\{{dM_n \over d\Omega dt}~{2 \over v_n}:n=1,2 ...\} =
(26,~11,~7,~5,~4,~...)10^{-2}~v_{rot}^2/4 \pi G$.  

The amount of angular momentum is related to the effective core radius 
of the galactic halo \cite{twi}.  For galaxies like our own, the average 
of the dimensionless angular momentum distribution $\bar j \simeq 0.2$.
For a given $j$ distribution, $\bar j$ and $j_{max}$ are related by 
some numerical factor.  If the turnaround sphere is initially rigidly
rotating, $j_{max} = {4 \over \pi} \bar j$.

If the caustic rings lie close to the galactic plane, they may manifest
themselves as bumps in the rotation curve.  Galactic rotation curves 
often do have bumps.  Of special interest here are those which occur
at radii larger than the disc radius because they cannot readily 
be attributed to inhomogeneities in the luminous matter distribution.  
Consider the rotation curve of NGC 3198 \cite{vA},  one of the best 
measured and often cited as providing compelling evidence for the
existence of dark halos.  It appears to have bumps near 28, 13.5 and 
9 kpc, assuming $h = 0.75$.  Although the statistical significance of 
these bumps is not great, let's assume for the moment that they are real 
effects.  Note then that their existence is inconsistent with the 
assumption that the dark halo is a perfect isothermal sphere.  On the 
other hand, the radii at which they occur are in close agreement with 
the ratios predicted by the self-similar model assuming that the bumps 
are caused by the gravitational fields of the first three caustic rings 
of NGC 3198.  Since $v_{rot} = 150$ km/sec, we find that $j_{max} = 0.28$ 
in this case if $\epsilon = 0.3$.  The uncertainty in $h$ drops out.  A 
fit of the infall model to our own galaxy \cite{twi} produced 
$\bar j \simeq 0.2$ for $\epsilon = 0.2$ to $0.3$.  If the turnaround 
sphere is initially rigidly rotating, one has 
$j_{max} = {4\over \pi}\bar j$.  Thus the values of $\bar j$ for our 
own halo and that of NGC 3198 are found to be similar.

Let's discuss the implications of the self-similar model for our own halo. 
As an example, we use the model parameters $\epsilon = 0.28$, 
$j_{max} = 0.25$ and $h=0.7$.  Table 1 shows the caustic ring radii 
$a_n$, the local velocities $\vec v_n$ and the local densities $d^{\pm}_n$, 
at the position of the Sun, associated with the first 20 pairs of flows. 
$\hat z$ is the direction perpendicular to the galactic plane, $\hat r$ is 
the radial direction in the galactic plane and $\hat \phi$ is the direction 
of galactic rotation.  The velocities are given in the rest frame of the 
Galaxy.  $d^{\pm}_n$ is the density of {\it each} of the two $n$th flows.
We at 8.5 kpc are between the 4th and 5th ring.  The local densities of the 
4th and 5th flows are large because of our proximity to the corresponding 
rings.

\begin{table}[t]
\caption{Caustic ring radii, local velocities and local densities of the 
first 20 pairs of flows in the self-similar infall model with 
$\epsilon = 0.28, j_{max} = 0.25$ and $h = 0.7$~.}
\vspace{0.2cm}
\begin{center}
\footnotesize
\begin{tabular}{|r|c|c|c|c|c|c|}
\hline
$n$ & $a_n$ & $v_n$ & $v_{n\phi}$ & $v_{nz}$ & $v_{nr}$ & $d_n^{\pm}$\\
& (kpc) & (km/s) & (km/s) & (km/s) & (km/s) & ($10^{-26}$ gr/cm$^3$)\\
\hline
1 & 38. & 620 & 140 & $\pm$605 & ~~0~~ & 0.4 \\
2 & 19. & 565 & 255 & $\pm$505 & ~~0~~ & 1.0 \\
3 & 13. & 530 & 350 & $\pm$390 & ~~0~~ & 2.0 \\
4 & 9.7 & 500 & 440 & $\pm$240 & ~~0~~ & 6.3 \\
5 & 7.8 & 480 & 440 & ~~0~~ & $\pm$190 & 9.2 \\
6 & 6.5 & 460 & 355 & ~~0~~ & $\pm$295 & 2.9 \\
7 & 5.6 & 445 & 290 & ~~0~~ & $\pm$330 & 1.9 \\
8 & 4.9 & 430 & 250 & ~~0~~ & $\pm$350 & 1.4 \\
9 & 4.4 & 415 & 215 & ~~0~~ & $\pm$355 & 1.1 \\
10 & 4.0 & 400 & 190 & ~~0~~& $\pm$355 & 1.0 \\
11 & 3.6 & 390 & 170 & ~~0~~& $\pm$355 & 0.9 \\
12 & 3.3 & 380 & 150 & ~~0~~& $\pm$350 & 0.8 \\
13 & 3.1 & 370 & 135 & ~~0~~& $\pm$345 & 0.7 \\
14 & 2.9 & 360 & 120 & ~~0~~& $\pm$340 & 0.6 \\
15 & 2.7 & 350 & 110 & ~~0~~& $\pm$330 & 0.6 \\
16 & 2.5 & 340 & 100 & ~~0~~& $\pm$325 & 0.55 \\
17 & 2.4 & 330 & ~90 & ~~0~~& $\pm$320 & 0.50 \\
18 & 2.2 & 320 & ~85 & ~~0~~& $\pm$310 & 0.50 \\
19 & 2.1 & 315 & ~80 & ~~0~~& $\pm$305 & 0.45 \\
20 & 2.0 & 310 & ~75 & ~~0~~& $\pm$300 & 0.45 \\
\hline
\end{tabular}
\end{center}
\end{table}

There is evidence for the 6th through 13th caustic rings in that there are 
sudden rises in the inner rotation curve \cite{Cl} of our galaxy at radii 
very near those listed in the Table.  Similarly there is some evidence for 
the 2d and 3d caustic rings in the averaged outer rotation curve \cite{Tre}.  
This will be discussed in detail in a future paper\cite{tbp}.  

This study was motivated by the axion \cite{ad} and WIMP \cite{sl} dark 
matter searches.  These experiments may some day measure the energy spectrum 
of CDM particles on Earth.  I would like to stress here the relevance of the 
model for the annual modulation of the signal in direct searches for WIMP dark 
matter.  The $n$ = 2, 3 ... 8 pairs of flows have velocities exceeding the 
220 km/s velocity of the Sun in the direction $\hat \phi$ of the Sun's 
motion.  A simple calculation \cite{tbp,Krau} shows that they dominate the 
annual modulation because of their large contributions to the local halo 
density.  Thus the annual modulation of the WIMP signal has the opposite 
sign in this model from that predicted by the isothermal sphere model.  

The model is also relevant to calculations \cite{Berg} of the sky-map of
gamma radiation from WIMP annihilation in the galactic halo.  The 
annihilation rate is proportional to the square of the density.  If the
diffuse gamma ray component seen by EGRET \cite{Dix} is attributed to
annihilations in the halo away from the disk, there ought to be hot spots 
in the disk at the location of the rings.

\section*{Acknowledgments}
I am grateful to K. Freese, B. Fuchs and L. Krauss for stimulating 
comments.  This work is supported in part by the U.S. Department of 
Energy under contract DE-FG05-86ER40272 .

\end{document}